\documentstyle[stwol]{article}

\input{psfig}

\begin{document}

\title       {CHAOS, SCALING AND EXISTENCE OF A CONTINUUM LIMIT \\
              IN CLASSICAL NON-ABELIAN LATTICE GAUGE THEORY}


\author      {Holger Bech Nielsen}
\address     { The Niels Bohr Institute, Blegdamsvej 17, 2100
       K\o benhavn \O, Denmark}
\author      {Hans Henrik Rugh}
\address     {Department of Mathematics, University of Warwick,
        Coventry, CV4 7AL, England}
\author      {Svend Erik Rugh}
\address     {Theoretical Division, T-6, MS B 288, University of
        California, Los Alamos National Laboratory, \\ Los Alamos,
         New Mexico, NM 87545, U.S.A.}


\twocolumn[\maketitle
\abstracts   {
We discuss space-time chaos and scaling properties
for classical non-Abelian gauge fields discretized
on a spatial lattice. We emphasize that there is a ``no go'' 
for simulating
the  original continuum classical gauge fields over a long time span 
since there is a never ending dynamical cascading towards
the ultraviolet. We note that the
temporal chaotic properties of the original continuum gauge fields and the
lattice gauge system have entirely different scaling properties
thereby emphasizing
 that they  
are entirely different dynamical systems 
which have only very little in common.
Considered as a statistical system in its own right
the lattice gauge system in a situation where it has reached
equilibrium comes  closest to
what could be termed a ``continuum limit'' in the limit of very
small energies (weak non-linearities).  We discuss the
lattice system both in the limit for small energies and in the
limit of high energies where we show that there is a saturation of the
temporal chaos as a pure lattice artifact. Our discussion
focuses
not only on the temporal
correlations but
to a large extent also on the spatial correlations in the lattice
system. We argue that various conclusions of physics
have been based on monitoring the non-Abelian lattice system in 
regimes where
the fields are correlated over few lattice units only. This is
further evidenced 
by comparison with results for Abelian lattice gauge
theory. How the real time simulations of the classical lattice gauge 
theory may reach contact with the real time evolution of 
(semi-classical aspects of) the quantum gauge theory
(e.g. Q.C.D.) is left as an important question to be 
further examined.   
}
]


\newcommand{\tr}   {\mbox{tr}}
\newcommand{\half} {\frac{1}{2}}
\newcommand{\r}    {\rightarrow}
\newcommand{\laB}{\lambda_B}
\newcommand{\PPsi}{{\bf X}}
\newcommand{\dP}{{\delta \bf X}}
\newcommand{\plaq}{{\Box}}
 
\newcommand{\simleq}{{\leq}}

There are some indications - and it would be a beautiful principle
if it was true - that we have the ``approximative laws and regularities''
which we know,
because they are  infrared stable against
modifications of ``short distance physics'' in the 
ultraviolet \cite{RandomDyn}.
These ``approximative laws'', which at present accessible
scales to the best of our knowledge comprise 
a sector consisting of the Standard Model of quantum Yang-Mills fields based
on the group $S(U_2 \times U_3) \sim U(1) \otimes SU(2) \otimes
SU(3)$ and a sector consisting of the gravitational interactions, 
may (thus) be robust and stable in the sense of Ref.\cite{RandomDyn},
but they are
full of unstable and chaotic solutions! 
Indeed, both the gravitational
field (the Einstein equations) 
and the Yang-Mills fields exhibit dynamical chaos 
\cite{Hobilletal,BiroetalBook}
for generic solutions
(solutions without too much symmetry) in regions 
where (semi)classical treatments are justified and
non-linearities of the interactions are non-negligible. In the present
contribution we shall report on some observations 
\cite{HBNetal96a,Muller96a,HBNetal96b}
concerning the lattice implementation of 
space-time chaos of classical Yang-Mills fields, 
but let us first note some 
few physical motivations for studying dynamical chaos in classical
Yang-Mills 
fields: 

(1) Despite the word ``chaos'' at first suggests something
structureless it is really a  
study of ways in which ``structure'' 
may be generated (from various initial configurations) 
during the evolution of the governing equations of motion. It is of
 interest - but so far only little is known - to know how structures
form and evolve (e.g. the formation and evolution of embedded
 topological structures)
during the time evolution
of the Yang-Mills equations. In the semi-classical regimes 
of the Standard Model, the quantum solutions build 
around the classical solutions
(and it is not unlikely 
that quantum fluctuations on top of
classical chaos only enhance chaos \footnote{The remark that the full
quantum theory for a system with bounded configuration space (and
a discrete spectrum of quantum states)  
has no chaos due to quasiperiodic evolution of the wave function
is of little importance to us since
the number of degrees of freedom involved here are so large that
the time it will take to evolve quasi-periodically through
the states will
be substantially larger than the lifetime of the Universe.}).

(2) For example, the possibility of baryon non-conservation 
(via the famous ABJ-anomaly) within the electroweak theory has achieved
quite some attention recently, and it is related to a detailed
understanding of the {\em dynamics} of the electroweak fields (as the
Universe cools down).
It is quite natural to speculate
about a relationship between dynamical chaos and an activity of
formation and destruction of embedded topologically interesting
field configurations (such a relationship is, for example, well known
for the complex Landau Ginzburg equation, see also e.g. discussion 
in Ref. \cite{RughYMlic94} which contemplates the relevance of a concept of
{\em topological turbulence} of the gauge fields).
We note that field configurations for which the rate of production
of baryon number 
\begin{equation} \label{ABJ}
\dot{B} \sim \int_{\Omega \subset R^3} \;  d^3 x \; 
Tr ( F \; F^{*})   
\end{equation}
vanishes will span a surface of co-dimension one. Thus, in fact,
most field configurations (in the
hot electroweak plasma) will 
contribute to the right hand side of equation (\ref{ABJ}).
For a particular class of field
configurations which contributes to the baryon non-conservation 
see also Ref.\cite{Axenidesetal95}.

(3) Fast equilibration processes which 
take place in heavy-ion collisions \cite{HotHadronicMatter95}
are very likely connected to non-linear chaotic dynamics; an idea which
in principle dates back e.g. to Fermi, Pasta and Ulam, Ref. \cite{FPU}.

As a further motivation of the study of {\em classical} Yang-Mills
chaos, we should also note that not many non-perturbative
tools are available to study the time evolution of quantum Yang-Mills
fields, so the study of classical Yang-Mills fields is a
natural starting point for
semi-classical understanding of the dynamics.
 
In order to facilitate a numerical study of spatio-temporal chaos 
(formation of space-time structure)
in inhomogeneous Yang-Mills fields 
one has 
in some way to discretize the space-time continuum on which the
Yang-Mills fields are defined. This can be
achieved in a way which breaks gauge invariance (see e.g. \cite{Wellner})
or in a gauge invariant way which is called
lattice gauge theory \cite{WilsonKogutSusskind}.
 
We shall refer to \cite{HBNetal96a,HBNetal96b} for more 
details in the discussion which will follow.
The study of chaos in toy-models where the gauge fields are
spatially homogeneous was initiated by Sergei Matinyan and
George Savvidy \cite{MatinyanSavvidy}.
An important new qualitative dynamical feature
comes into play, however, when one considers 
the spatially inhomogeneous classical
Yang-Mills equations: There is a never ending cascading of the dynamical
degrees of freedom towards the ultraviolet, generated by the time evolution
of the Yang-Mills equations!
(In spite of this ``ultraviolet catastrophe'', the solutions are
well behaved in the sense 
that there are
no ``finite time blow up of 
singularities''). 
The non-linear self-coupling terms which open up the
possibility for a chaotic behavior in the classical evolution
lead in the non-homogeneous case 
to the infinite cascade of energy
from the long wavelength modes towards the ultraviolet.
(Note, there is a priori no concept of temperature
and in the case of Abelian (electromagnetic) fields
this cascading would 
\underline{not} show up
dynamically, unless
one couples the fields to charged particles).
This tendency of the mode frequencies cascading
towards the ultraviolet will
completely dominate the qualitative behavior of the 
{\em classical} Yang-Mills
equations, and the ``ultraviolet catastrophe'' has for some time been
emphasized by us (cf. e.g. discussion in Ref.\cite{RughYMlic94})
as a major obstacle to simulate  
the classical continuum Yang-Mills fields in a numerical experiment
over a long time span. \footnote{This is an
obstacle which, in our opinion, has received
insufficient  attention in the various studies 
attempting at discussing and modeling
chaotic properties of spatially inhomogeneous classical Yang-Mills fields.}
There is no mechanism, within the classical equations, which prevents this
never ending cascading of the modes towards the ultraviolet.
Nature needs $\hbar$, the Planck constant, as an ultraviolet regulator.
Indeed, both Abelian and non-Abelian gauge fields are implemented as quantum 
theories in Nature.

Here we shall 
discuss the possibility 
of using a lattice cutoff in a purely
classical treatment to regularize the equations.
There will still be a cascading of modes towards the ultraviolet, i.e. towards
the lattice cut-off, and this ultraviolet cascade will still dominate the
dynamical evolution of smooth initial field configurations.
However, in a lattice formulation of the Yang-Mills fields 
on a large but finite lattice,
the phase space is compact  for any given energy and thus
the system  can reach an equilibrium state among the modes 
(a `thermodynamic equilibrium').
The lattice regularization of the theory opens up for 
the definition of
dynamic and thermodynamic properties,
which are not defined in the classical Yang-Mills field
theory without regularization. It could e.g. be
ergodic (modulo constraints) with respect to the Liouville measure,
in which case it makes 
sense to talk about its micro-canonical distribution
and approximating this by looking at 
`typical' classical trajectories. The fundamental
assumption of thermodynamics asserts that 
on average the two 
approaches give the same result if we have a large system, 
and we may then naturally introduce  correlation functions and 
possibly a correlation length $\xi$ (measured in lattice units)
of the system.

 One hopes to
define a continuum theory if, 
by judicious choice of the parameters in the system,
one obtains a
physical correlation length in the limit when the lattice
constant goes to zero, i.e.
\begin{equation} \label{finitecorr}
\xi(a, E(a),...) \times a  \; \; \r
\; \; \ell  \neq 0 \; \; , \; \; \mbox{as} \ a\r 0.
\end{equation}
In equation (\ref{finitecorr}) the 
correlation length $\xi$ in the
lattice system is a function of lattice model parameters such as 
lattice spacing $a$, average energy density $E(a)$, etc.
Condition (\ref{finitecorr}) implies
that the correlation length diverges when measured in lattice units
and only if this is the case do
we expect the lattice
system to lose its memory of the underlying lattice structure.

We shall restrict attention to
the lattice gauge theory in
3+1 dimensions based on the gauge
group $SU(2)$. We consider a finite size
3 dimensional hypercubic lattice having $N^{3}$ points
where nearest neighbor 
points are separated
by a distance $a>0$.
The phase space 
is a fibered space where the tangent manifold of
the Lie-group $SU(2)$
is assigned to each of the links, $i \in \Lambda$,
connecting nearest neighbor lattice points: To each
$i \in \Lambda$ is associated a link variable
$U_i \in SU(2)$ as well as its canonical momentum\footnote
    {In principle $P_i$ is a cotangent vector, but the Lie algebra
     inner product gives a natural identification of the cotangent
     space with the tangent space.}
$P_i \in T_{U_i} SU(2)$. A point in the entire phase space
 will be denoted
\begin{equation}
\PPsi = \{U_i, P_i\}_{i \in \Lambda} \ \in \ 
M = \prod_\Lambda T \; SU(2)\ .
\end{equation}
A Hamiltonian is constructed so as to correspond to the
continuum, classical Yang-Mills Hamiltonian in the limit
$a \rightarrow 0$. The construction of such a Hamiltonian 
is of course ambiguous in the sense that extra terms of order
$O(a)$ may be added to the Hamiltonian. Alluding to some sort of
``universality''
(especially in the limit $a \rightarrow 0$), 
we expect that the precise choice of Hamiltonian is not so
important in what follows.
A Hamiltonian which is often employed to  
generate the time evolution of the orbit $\PPsi (t)$
is the Kogut-Susskind Hamiltonian, which 
can be written in the following way \footnote{For a derivation
we refer to \cite{WilsonKogutSusskind} and  
\cite{BiroetalBook,Muller} from
which we adapt our notation.
For simplicity
we omit the coupling constant factor $2/g^2$ which anyway is arbitrary in
a classical theory.}~:
\begin{eqnarray} \label{HamiltonKS}
  \lefteqn{H(a, \PPsi^{(a)}) =} \nonumber \\
   &&
      \frac{1}{a}
       \sum_{i \in \Lambda} \half \tr (P_i P_i^\dagger) +
      \frac{1}{a}
   \sum_\plaq(1  - \half \tr U_{\plaq})  \ .
 \label{eq:kogut}
\end{eqnarray}
Here the last sum is over elementary plaquettes
bounded by 4 links, and $U_{\plaq}$ denotes the
path-ordered product of the 4 gauge elements along the
boundary of the plaquette ${\plaq}$.
The last term, the potential term, is automatically bounded and, 
for a given finite total energy, the same is the case for 
the first term, the kinetic term. Thus the phase space
corresponding to a given energy-surface is compact.

As is standard  we shall discuss the 
temporal correlations (temporal
chaos) of the lattice gauge system in terms of its spectrum of Lyapunov 
exponents. The compactness of the phase space implies 
\cite{HBNetal96a}
that the spectrum of
Lyapunov exponents (which we overall will assume to be well defined 
quantities for the lattice system) is independent of the
choice of norm on the
space of field configurations. 
In fact, it follows \cite{BiroetalBook,HBNetal96a} 
from the scaling properties of the equations of motion
generated by the lattice Hamiltonian (\ref{HamiltonKS}),
that in order to study the dependence of the maximal Lyapunov
exponent with the energy density of the system, it is sufficient to consider
the equations of motion for a fixed value of  the lattice constant
$a$, e.g. $a=1$,  as a function of energy density
($\propto$ energy/plaquette) and
then rescale  the results back afterwards.

We have several different forms of lattice artifacts in the lattice 
simulation of real-time dynamical behavior of the continuum classical
 Yang-Mills fields:

(1) Lattice artifacts due to the compactness of the group.
 The magnetic term (the second term) in the Kogut-Susskind
Hamiltonian (\ref{HamiltonKS}) is uniformly bounded,
$ 0 \leq 1 - \frac{1}{2} Tr U_{\Box} \leq 2$, due 
to the $SU(2)$ compactification. Thinking in terms of statistical 
mechanics for
our classical lattice system, we expect that after some time the 
typical field configuration has equally much energy in all modes
of vibration - independent of the frequency\footnote{Note, this situation is
very different in the quantum case. 
Planck's constant $\hbar$ introduces a relation
$E = \hbar \omega$ between the energy of a mode of vibration
and its frequency, 
implying that a mode with a high frequency also has a high energy.
With a given available finite total energy,  modes with
high frequencies will therefore be suppressed. 
Quantum mechanically, we thus
have that at low energy only excitations of the longest
wavelengths appear.} - and the total
amplitude of the classical field, and the energy per lattice
plaquette, is thus small for a fixed low energy. For low energy,
when the average energy per plaquette is small, the 
lattice artifacts due to the compactness of the gauge
group are thus negligible.  

(2) For small energy per plaquette, we thus expect that the dominant 
form for lattice artifacts is due to the fact
that an appreciable amount
of the activity (for example the energy) is in the field modes
with wavelengths comparable to the lattice constant $a$.
This short wavelength activity at lattice cut-off scales
is unavoidable in the limit of long time simulation of
an initially smooth field configuration (relative
to the lattice spacing),
or already after a short time if we initially have an irregular
field configuration.

In which way can the classical lattice gauge theory approach a {\em continuum
limit} ? This is a difficult question to which we shall only be able
to provide a very partial answer here.

For the study of the time evolution of Yang-Mills fields
which initially are {\em far from 
an equilibrium situation}, the (classical)
field modes will exhibit a never ending dynamical cascade towards the 
ultraviolet and after a 
certain transient time, the cut-off
provided by the spatial lattice will prevent the lattice gauge theory
from simulating this cascade.
It is therefore immediately clear that
the lattice regularized, classical fields will not approach a
``continuum limit'' in the sense of simulating the dynamical
behavior of the classical {\em continuum} fields in the $t \rightarrow \infty$
limit.
For the simulation of classical continuum gauge fields far from 
equilibrium, we conclude that we will have
the best ``continuum limit'' if we simulate,
for a short period of time, an initial
smooth ansatz\footnote{By ``smooth'' configurations
we mean lattice configurations which are `good' approximations
to continuum configurations.} for the fields in the region
of low energy per plaquette.\footnote{I.e. we are here imagining a 
situation where the spatial correlations in the monitored
field variables are so large that they
lose memory of the underlying lattice structure (including  the lattice
spacing $a$). In the extreme opposite limit,
one could imagine situations with randomly
fluctuating fields on the scale of the lattice constant,
i.e. with (almost) no spatial correlations from link to link.
If the field variables fluctuate independently of each other
(independent of their neighbors), one could imagine the
model to be invariant (with respect to the monitoring of many
variables) under changes of
the lattice spacing, $a$. Thus, it appears that
lattice cut-off independence of numerical results can
\underline{not} be
a sufficient criterion for the results to report ``continuum physics''.}

Gauge fields exist, however, as quantum 
theories in Nature and the interesting definition - as concerns
applications in {\em physics} - of 
a ``continuum limit'' of the classical
lattice gauge theory, is thus to
identify  regions in the parameter space for the classical
lattice gauge theory which
 probe the behavior of the time evolution of semi-classical
initial configurations (with many quanta) of the quantum
theory for a shorter or longer interval of time or in
an equilibrium situation. (See also e.g. discussion in Ref.\cite{Bodekeretal}).
Since the classical lattice gauge theory does not contain 
a relationship like $E = \hbar \omega$ (implying a damping of the high
frequency modes relative to the soft modes), we must expect that
the simulation of quantum gauge theory will be distorted by this
fact (even if implemented with effective lattice Hamiltonians as
e.g. devised by Ref.\cite{Bodekeretal}).

We shall in the following restrict attention to
real-time simulations of the classical lattice 
gauge theory which have reached an {\em equilibrium situation}
on the lattice. Such simulations are hoped
\cite{BiroetalBook} to yield
insights into the dynamical behavior of 
(long wavelength modes in) the high temperature
Q.C.D. fields in situations where an equilibrium situation
has been reached. 

Before we attempt an analysis of aspects of chaos (in time and space)
of the lattice gauge theory let us first
note that it is far from obvious how to 
establish contact between 
an {\em effective lattice temperature} of
the {\em classical} lattice gauge theory (which has reached an equilibrium
situation due to the presence 
of the lattice cutoff
$\Lambda = 1/a$)
and the {\em physical} temperature $T$ of the Q.C.D. theory
which is a {\em quantum} theory and which 
can reach an equilibrium situation due to a cut off of
quantum mechanical origin (i.e. ultimately due to the
existence of a Planck constant $\hbar$).
Some discrepancy in the literature
\cite{Ambjorn,Biroetal96,Heinzetal96}
illustrates that this is not
an easy question.

In the classical lattice gauge theory
we have an ``effective lattice temperature'' $T_{L} = 1/\beta$ where
$\beta$ has been determined by looking at the probability
distribution (Gibbs distribution) of the kinetic energy (or the
magnetic energy in a plaquette),  
$p(E_k) \propto \exp (-\beta E_k)$ (after equilibration). Cf. e.g. Ref.
\cite{BiroetalBook} (p. 206-207) and Ref. \cite{Heinzetal96}.

In Ref.\cite{BiroetalBook,Biro,Heinzetal96}
it is asserted that the
physical temperature $T$ (of the quantum Q.C.D. fields 
in an equilibrium situation)
and the average energy $E_{\Box}$ per plaquette
in the classical lattice gauge study in equilibrium are related by
\begin{equation} \label{temp}
E_{\Box}  \approx \frac{2}{3} \; (n^2 - 1) \;  T  \; \;  \;
\mbox{(for} \; SU(n) \mbox{)}.
\end{equation}
It is not clear to us how serious one should take this relationship
(one objection being that a scale has not been fixed in the continuum limit
 of the classical theory) but we shall leave a discussion of this
issue aside in the following.

Let us now attempt an understanding of chaotic aspects of the lattice
gauge theory - i.e. the temporal chaos (as monitored by the spectrum
of Lyapunov exponents) and its relationship with spatial chaos
(as monitored by the spatial correlations, e.g. a spatial correlation
length) in the dynamical system. 

A sequence of articles\cite{Muller,Gong,Biro}
and a recent book 
by Bir\'{o}, Matinyan and M\"{u}ller\cite{BiroetalBook}  present
 numerical results for the classical $SU(2)$ lattice gauge model
and provide evidence
that the maximal Lyapunov exponent is
a  monotonically
increasing continuous function of the scale free energy/plaquette
with the value zero at zero energy.

Ref.\cite{BiroetalBook,Muller,Biro} reports a particularly interesting
interpretation of numerical results for the dynamics on the lattice,
namely that there is
a {\em linear} scaling relation between the scale free maximal Lyapunov 
exponent,
$\lambda_{max}(a=1)$ and the  average energy per plaquette 
$E_{\Box}   (a=1)$.
The possible physical relevance of this result
follows from the observation\cite{Muller} that when we
rescale back to a variable lattice spacing $a$ 
we note that
the observed relationship is in fact a graph of
$a \lambda_{max}(a)$ as a function of $a \; E_{\Box}(a)$. Thus being
linear, 
\begin{equation} \label{linearlambda}
a \; \lambda_{max} = \mbox{const} \times a \; E_{\Box} (a)
\end{equation}
cancellation of a factor $a$ implies that 
$\lambda_{max}(a) = \mbox{const} \times E_{\Box} (a)$ 
and thus there is
a continuum limit $a\r 0$, either of both sides simultaneously
or of none of them. In the particular case where the energy
per mode ($\propto$ energy per plaquette) is interpreted to be a
fixed temperature $T$ (cf. equation (\ref{temp}) above), 
one deduces that the maximal
Lyapunov exponent has a continuum limit in real time, proportional
to the temperature of the gauge field.

Note, also, the suggested relationship\cite{BiroetalBook,Muller,Biroetal95} 
between the maximal Lyapunov
exponent $\lambda$ of the gauge fields
on the lattice and the ``gluon damping rate''
$ \gamma (0)$  for a thermal gluon at rest,
arrived at in re-summed perturbation theory in finite temperature
quantum field theory, see also Refs.\cite{Biroetal96,Heinzetal96}.
For the $SU(2)$ gauge theory, this suggested relation reads
\begin{equation} \label{plasmon}
\lambda = 2 \gamma (0) = 
2 \times 6.635 \frac{2}{24 \pi} g^2 T \sim
0.34 \; g^2 T 
\end{equation}
It is also our understanding that 
$\gamma (0)$ is a quantity of semiclassical origin.\cite{Gong}
A relation like (\ref{plasmon})  is nevertheless remarkable in
chaos theory since it suggests that
a complicated dynamical quantity like a temporal Lyapunov
exponent (which is usually only possible
to extract after a considerable numerical effort)
is analytically calculable by summing up 
some diagrams in finite temperature quantum
field theory.

In Ref.\cite{HBNetal96a} we argued (in view of
the numerical evidence presented in e.g. 
\cite{BiroetalBook,Muller,Biro})
that the apparent linear scaling relation (\ref{linearlambda})
is a transient phenomena residing in a
region extending at most a decade
 between two scaling regions, namely 
for small energies 
where the Lyapunov exponent scales with an
exponent which could be close to $k \sim 1/4$, 
and a high energy region where
the scaling exponent is at most zero:
\begin{equation}
\eta = \frac{d \log \lambda}{d \log E} =
\left\{  \begin{array}{ll}
= k \sim 1/4  & \mbox{for $E \r 0$}  \\
\leq 0 & \mbox{for $E \r \infty$}
\end{array}
\right.
\label{eq:our}
\end{equation}
The proposed scaling relation $k \sim 1/4$
was, for reasons we shall give below, suggested to hold in the limit
$E \rightarrow 0$ 
from general scaling arguments 
\cite{RughYMlic94} of the
continuum classical Yang-Mills equations in accordance with simulations on
homogeneous models and consistent with the figures in
\cite{Muller} and \cite{Biro}.
However, a more recent numerical analysis \cite{Muller96a}
 argues rather convincingly that the data points
presented in e.g. \cite{Muller,Biro,BiroetalBook} were subject to 
``finite time''
artifacts, and that long time simulations with
a correct procedure for extracting the principal Lyapunov
exponent could support
that $k \equiv 1$ even in the limit as $E \rightarrow 0$.

Before we discuss the limit of
low energies ($E \rightarrow 0$)
let us note that regarding 
 the limit for high energies per plaquette a rigorous result
\cite{hhrugh} (also reported in \cite{HBNetal96a,HBNetal96b}) 
shows that the $SU(2)$ scale free ($a=1$)
 lattice Hamiltonian in $d$
spatial dimensions
 has an upper bound
for
the maximal Lyapunov exponent
\begin{equation} \label{laB}
 \laB = \sqrt{(d - 1) (4+\sqrt{17} )}
\end{equation}
which for $d=3$  becomes $\laB = 4.03..$. 
This result  is arrived at by constructing
an appropriate  norm on the phase
space 
and showing that the time derivative of this norm can be bounded
by a constant times 
 the norm itself, hence giving us an upper bound as to how 
exponentially fast the
norm can grow in time. 
The upper bound (\ref{laB}) shows that a linear scaling
region, i.e. $\eta = d \log \lambda / d \log E = 1$,  
cannot extend further than around $E \sim 10$
on the
figure 1.
Beyond that point the maximal Lyapunov
exponent either saturates and scales with energy with an exponent which
approaches zero or it may even decrease
over a region of high energies, yielding a negative
 exponent thus justifying the
$E \rightarrow \infty$ limit of equation (\ref{eq:our}).

It should be noted that the upper bound 
(\ref{laB}) is independent of the lattice size and the energy
(but scales with $1/a$).
There is quite a simple intuitive explanation for the
saturation of the maximal Lyapunov exponent in the regime for
high  energy per plaquette of the 
 $SU(2)$ lattice gauge model, since 
the potential (magnetic) term in the Hamiltonian (\ref{eq:kogut})
is uniformly bounded,
$ 0 \leq 1 - \half  Tr U_{\Box} \leq 2$, due to the $SU(2)$ compactification.
For high energies, almost all the energy is thus put in the 
(integrable) kinetic energy term in the 
lattice Hamiltonian (\ref{eq:kogut}), and the
spectrum of Lyapunov exponents will saturate as the energy per plaquette
increases for the model. (The chaos generated by the non-linear
potential energy term
does not increase, only the energy in the kinetic energy
(the electric fields) increases).

As we shall argue (a discussion which
is rather suppressed in \cite{Muller,Gong,Biro,BiroetalBook}),
it is however in the opposite limit,  
i.e. in the limit where the average energy per plaquette 
goes to zero, i.e. $ E \rightarrow 0$,
that the equilibrium lattice gauge theory has 
the possibility to
reach contact with ``continuum physics'' -
\underline{not} in the sense (as
 we have seen) that the lattice system simulates
the original continuum classical gauge fields (over a long
time span), but in the sense that the lattice
gauge theory - considered as a statistical system in its
own right - will develop spatial correlations in the fields
(as monitored by a correlation length $\xi$, say) which may
approach a formal ``continuum limit'',
\begin{equation}
\xi \rightarrow \infty  \; \; \mbox{as} \; \; E \rightarrow 0 
\end{equation}
There is a well known analogy\cite{analogy}
between a quantum field theory in the Euclidean
formulation with a compactified (periodic) imaginary time 
axis $0 \leq \tau \leq \beta$ 
and finite temperature statistical mechanics of the quantum
field theory at a temperature $ T = 1/\beta$
which is inversely proportional to the above-mentioned  time extension.

Calculations in finite temperature Euclidean 
quantum field theory 
suggest that a characteristic correlation length of static magnetic
fields in the thermal quantum gauge theory
is of the order $\xi \sim (g^2 T)^{-1}$. If we (cf. discussion above,
equation (\ref{temp}))
make use of the relationship, $  E_{\Box}   = 2 T$ (for 
SU(2) Yang-Mills theory) between temperature $T$ and average 
energy per plaquette 
$ E_{\Box} $, this could suggest
that the (magnetic sector) of field configurations on the lattice
will have characteristic correlation lengths of the order
\begin{equation} \label{correlationT}
\xi \sim (g^2 T)^{-1} \sim ( \frac{1}{2}  g^2  E_{\Box}  )^{-1} 
\rightarrow \infty \; \; \mbox{as} \;  \; E_{\Box} \rightarrow 0 \; .
\end{equation}
In systems with space-time chaos (and local
propagation of disturbances with a fixed speed $\sim c$)
one often has a relationship between a
correlation length (coherence length) $\xi$ and the
maximal Lyapunov exponent $\lambda$ (see also e.g. discussions 
in Ref.\cite{TBohr}) of the form $\xi \sim c / \lambda$.
This suggests, via equation (\ref{correlationT}),
\begin{equation}
\lambda \sim c/ \xi \sim  \; g^2 \; E_{\Box}
\end{equation}
i.e. a Lyapunov exponent which scales linearly with the
average energy per plaquette $E_{\Box}$.
Clearly, it deserves further investigation to establish 
a precise relationship between $\lambda_{max}$ 
and the spatial correlation length $\xi$ (in lattice units).
However, we emphasize that it is in the limit as the energy per plaquette 
goes to zero 
(and where the lattice Lyapunov exponent
$\lambda_{max}$ goes to zero)
that we expect  the spatial correlation length $\xi$ 
(in lattice units) will diverge.
The finite size of the lattice makes
it difficult to 
analyze the behavior of the gauge fields and the principal Lyapunov
exponent on the lattice in the limit $E \rightarrow 0$.

Let us reiterate the arguments which led us to the conclusion
(in Ref.\cite{HBNetal96a}) that \marginpar{SER}
the scaling exponent of the Lyapunov exponent likely will be closer to
$\sim 1/4$ in the limit $E \rightarrow 0$
than the scaling exponent $\sim 1$ observed in the
intermediate energy region $1/2 \; \simleq E \; \simleq 4$.

In figure 1  we  display a log-log plot of the
results obtained in Ref.\cite{BiroetalBook,Muller,Biro}. It appears
that for $E \sim 1/2$ there is a cross-over to
another scaling region. Although the data points in this
region are determined with some numerical
uncertainty (finite time artifacts, as argued in Ref.\cite{Muller96a})
we did note that they are consistent with a
scaling with exponent $\sim 1/4$. 
Moreover,
it is well known that the homogeneous Yang-Mills equations 
have a non-zero
Lyapunov exponent
which by elementary scaling arguments
scales with the fourth root of the energy density and has the
approximate form \cite{Chirikov}:
\begin{equation} \label{ChirikovFormula}
\lambda_{\rm max} \approx  1/3 \; E^{1/4} \ .
\end{equation}
On the lattice  the above scaling relation is valid for spatially homogeneous
fields, i.e. the maximal Lyapunov exponent scales with the fourth root of the
energy per plaquette.
By continuity, fields
which are almost homogeneous \footnote{Such field
configurations are (ungeneric) examples of field configurations
which exhibit correlation lengths much larger than the lattice spacing.} 
on the lattice
will, in their transient, initial dynamical behavior,
exhibit a scaling exponent close to $1/4$.
In fact, the same  scaling exponent would also hold for
the inhomogeneous Yang-Mills equations \cite{RughYMlic94} had
the fields been smooth relative to the lattice scale,
so derivatives, $\partial_\mu$, are well approximated by their
lattice equivalent and
we are allowed to scale lengths
as well. This is seen from scaling arguments for the continuum Yang-Mills
equations
$ D_\mu F^{\mu \nu} = 0 $
 where
    $ D_\mu = \partial_\mu - i g [A_\mu, \; \; ] $
and  $\ F^{\mu \nu} = 
      \partial^\mu A^\nu - \partial^\nu A^\mu - ig[A^\mu,A^\nu]$.
Not taking boundary conditions into account,
these equations are invariant when $\partial_\mu$ and $A_\mu$ are
scaled with the same factor $\alpha$. That is,
if $A(x,t)$ is
a solution to the equations, then 
$\frac{1}{\alpha} A(\alpha x, \alpha t)$ is also a solution. 
The energy density $E$,
which is quadratic in the Yang-Mills
field curvature tensor $F^{\mu \nu}$,
then scales with $\alpha^4$.
This indicates \cite{HBNetal96a} that if we perform 
a measurement of the maximal
Lyapunov exponent $\lambda_{max}$ over a time short enough  for
the solutions to stay smooth, then $\lambda_{max}$ scales with
$E^{1/4}$. Such a scaling was also observed for ``smooth''
configurations by 
M\"{u}ller and Trayanov 
p. 3389 in Ref.\cite{Muller}. These scaling arguments
do not carry over to infinite time averages since solutions on
the lattice tend to be irregular. 

In any case we expect
that the lattice gauge theory
does not probe field configurations which are substantially
correlated beyond one lattice unit when the 
lattice gauge theory is monitored for energies per plaquette
in the regime $ E \sim 1/2 - 4 $.
That is, we expect
$ \xi / a  \sim 1  \; \; \mbox{for} \; \;
E \sim 1/2 - 4 $
and thus the system is indeed
very far from reaching contact with ``continuum
physics'' when monitored in that regime of energies. 
For example, the numerical results for the Lyapunov exponents as presented
e.g. in Ref. \cite{BiroetalBook,Muller,Biro} (c.f. e.g. fig. 8.4 in
Ref.\cite{BiroetalBook})
report on lattice size $N = 20$ for energies $E \sim 1/2 - 4$
and coupling strength $ g \approx 2$, which suggest
(cf. also the equation (\ref{correlationT})) that the lattice
gauge theory in this regime of parameters  monitor lattice
field configurations which have a correlation length up to
a few lattice units only. 
There is also numerical evidence \footnote{J. Ambj\o rn,
priv.communication. See also e.g. Ref. \cite{Kanaya} which
reports investigations of finite temperature QCD on the lattice
with measurements of a magnetic mass (inverse correlation length)
of the order $m_{mag} \sim 0.5 \; g^2 \; T$.}
that the  correlation
length for energies $E \sim 1$ is of the order of 
a few lattice units. This observation 
applies also to the numerical studies of the $SU(3)$ lattice gauge
theory reported in \cite{Gong,Biro} in the range of
energies per plaquette $E_{\Box} \sim 4-6$.

These observations are further substantiated by
the numerical simulations of Ref.\cite{Biro} (figure 12)
for the lattice gauge theory with a $U(1)$ group showing
a steep increase of the maximal Lyapunov exponent
with energy/plaquette in the 
interval $1 \; \simleq \; E \; \simleq \; 4$. 
The continuum
theory here
corresponds to the classical electromagnetic fields which have no
self-interaction, and thus the Lyapunov exponent in this limit
should vanish.
 The discrepancies in this case were 
attributed in \cite{Biro}
to a combined effect of the discreteness of the lattice and
the compactness of the gauge group $U(1)$  and were not
connected with finite size effects. This suggests strongly
that we, in the case of
numerical studies of a $SU(2)$ Kogut-Susskind Hamiltonian system, 
cannot base
continuum physics on results from simulations in the same interval of
energies where the $U(1)$ simulations fail to display continuum
physics.
On the contrary, we suspect that 
what could reasonably be called ``continuum physics'' 
(for the lattice gauge theory considered as a statistical system
in its own right) has to be extracted from
investigations of the Kogut-Susskind lattice simulations for energies 
per plaquette which are
at least much smaller than $ E \sim 1/2$. 

\begin      {figure}
\centerline {\psfig{file=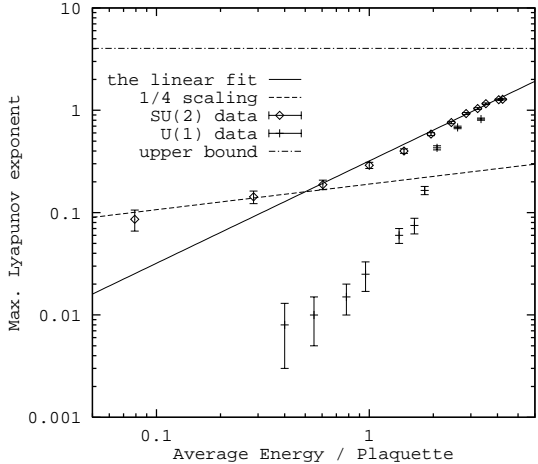,width=8cm}}
\caption    [Fig 1]
            {The maximal Lyapunov exponent as a function of
the average energy per plaquette for the $SU(2)$ and the $U(1)$
lattice gauge theory. 
The data points (diamonds for $SU(2)$, ticks for $U(1)$)
 are adapted from M\"{u}ller et al.
\cite{Muller} and Bir\'{o} et al. \cite{BiroetalBook}, p. 192.
The solid line is a linear fit
through the origin, the dashed line is the function
$\half \times 1/3 \times E^{1/4}$ (half of the homogeneous case result
(\ref{ChirikovFormula})).
The dot-dashed line
shows the rigorous upper bound for the $SU(2)$ lattice model
(saturation of temporal chaos).
As is seen, the linear scaling region for the $SU(2)$
data is positioned where the $U(1)$ data 
display strong lattice artifacts.
}
\label      {fig1}
\end        {figure}

We would like to conclude with some final points
of discussion.

Exactly in which way we may speak about a ``continuum limit'' in a
simulation of a classical gauge theory in real time on a 
spatial lattice (a theory which
does not have a continuum limit without lattice regularization)
appears to be a question of some fundamental as well as practical
interest, since not many non-perturbative tools are available to
study the time evolution of quantized Yang-Mills fields. 

Since by intrinsic 
scaling arguments (cf. Refs, \cite{BiroetalBook,HBNetal96a}) 
one has a functional relation between
$a\lambda_{max}(a)$ and $a E_{\Box} (a)$, a scaling (at small energies
$E \rightarrow 0$) with exponent
$k$, according to the scaling relation (\ref{eq:our}),
would imply that
\begin{equation}
\lambda(a) \propto a^{k - 1} E_{\Box} (a)^{k}  \; ,
\end{equation} 
in which case one cannot
achieve a continuum limit simultaneously for the maximal
Lyapunov exponent and the temperature (assuming that
it is proportional to the average energy per plaquette
 $E(a)$, as given by equation (\ref{temp})), except in the special
case where $k \equiv 1$. In particular,
if $k < 1$, the former would be divergent if the temperature is kept fixed.
There is no particular contradiction in this statement, however,
as there is - a priori - no 
 reason for having a finite
Lyapunov exponent in the continuum limit. The erratic and fluctuating behavior
of the fields 
one expects in time as well as in space 
(for numerical evidence, cf. also \cite{Wellner}) on very small
scales could suggest  that a Lyapunov exponent would not
be well defined in the ``continuum limit'' (as $a \rightarrow 0$ and
$E \rightarrow 0$). Clearly, this question deserves further investigation.

If the scaling-relation $k \equiv 1$ according to equation (\ref{eq:our})  
holds in the limit $E \rightarrow 0$ for the lattice
gauge theory, it will in a most striking
way illustrate the
point that the continuum gauge theory 
(with scaling $k \sim 1/4$) and the lattice gauge theory
(probed in a situation where it has reached
equilibrium) 
are two {\em entirely different} dynamical systems - despite  
the lattice theory (\ref{HamiltonKS})
is at first set up to be an approximation
to the continuum theory. (Thus the lattice theory does 
\underline{not} simulate the continuum theory; it is an entirely new 
statistical theory in its
own right and - important for applications in physics -
the relationship with quantum non-Abelian gauge
theory remains to be established on a more rigorous basis).
Understanding of this crucial difference could, perhaps, 
be obtained from renormalization group analysis: We
may say that the limit $E \rightarrow 0$ is a critical point for the
lattice Hamiltonian (\ref{HamiltonKS}),
 i.e. that the lattice correlation length 
   diverges in that limit. As is common for field theories studied in a
   neighborhood of a critical point, one expects classical scaling arguments
   to break down, or rather, that scaling relations are subjected to 
   renormalization which gives rise to anomalies in the scaling exponents.
   Often such anomalous scaling exponents seem to be 'ugly' irrational numbers,
   perhaps with the 2-d Ising model as a notable exception. It would therefore
   be quite miraculous if the $1/4$ classical-scaling of the Lyapunov
   exponent emphasized above renormalizes to an exponent which equals unity.
   Even if this did happen it is not clear that such a behavior should
   be independent of the regularization procedure employed, since Lyapunov
   exponents are measures of local instabilities, i.e. short range
   rather than long range structures.

As regards the numerical evidence for 
the relationship (\ref{plasmon}) between the
maximal Lyapunov exponent and the ``gluon damping rate''
\cite{Biroetal95} we note that in the hierarchy of scales
$ g^2 T \; \ll \;  gT \;  \ll  T  $ 
(this separation of scales is assumed in hot perturbative gauge theory)
the ``gluon damping rate'' is connected to the decay of the
``soft'' modes $\sim g^2 T$. In order for the lattice gauge theory
to probe decays of ``soft'' gauge modes, this requires 
the existence of some ``soft'' modes on the lattice (in a background
of ``hard'' modes). Thus we must monitor the lattice gauge 
system in a region where
there are fields (in equilibrium with ``hard'' modes $\Lambda \sim 1/a$)
which have spatial correlations substantially
larger than the lattice unit $a.$ 
As we have seen this is not the case in the regime where the
lattice gauge theory was studied ($N = 20, g \approx 2, E \sim 1 - 4$)
in Ref. \cite{BiroetalBook,Muller,Biro}
providing the numerical support
for equation (\ref{plasmon}).
It appears, that the numerical support of the
relation (\ref{plasmon}) is somewhat of an accident if the
relation (\ref{plasmon}) is to be interpreted as
a ``continuum result''.
Studies of the lattice gauge theory have to be conducted
for much smaller energies per plaquette which are, however, also
difficult, since finite size effects then will become of appreciable
size. \footnote{Finite size effects are of course -
in the context of our discussion - a ``good sign''
which witnesses that the lattice monitors field configurations
which have correlations of the order of the lattice size $\sim N a$, i.e.
much beyond a single lattice unit.}
\section*{Acknowledgements} 

We thank Alexander Krasnitz,
Salman Habib and Emil Mottola
for several discussions on the subject and
we thank Berndt M\"{u}ller for sending us
numerical data from Ref. \cite{Muller} and illuminating 
discussions following
the submission of our initial manuscript 
Ref. \cite{HBNetal96a}. 
S.E.R. would like to thank
the U.S. Department of Energy and Wojciech H. Zurek at the Los Alamos National
Laboratory for support. Support from Grant No. PHY94-07194
(National Science Foundation) from University of California 
Santa Barbara and discussions with Hans-Thomas Elze are also acknowledged.

%


\section*{References}
\begin{thebibliography}{99}

 
\bibitem{RandomDyn}
H.B. Nielsen, ``Dual Strings - Section 6. Catastrophe Theory
Programme'', in I.M. Barbour and A.T. Davies (eds.),
{\em Fundamentals of Quark Models}, Scottish Univ. Summer
School in Phys. (1976) pp. 528-543. 
See also (e.g.) reprints and discussion in 
C.D. Froggatt and H.B. Nielsen, {\em Origin of Symmetries}.
World Scientific. 1991.


\bibitem{Hobilletal}
D. Hobill et al. (eds.), {\em Deterministic Chaos in General Relativity},
NATO ASI Series B; Physics Vol. {\bf 332}. Plenum Press. New York. 1994.
 

 
\bibitem{BiroetalBook}
T.S. Biro, S.G. Matinyan and B. M\"{u}ller,
{\em Chaos and Gauge Field Theory}. World Scientific. 1994.
 



\bibitem{HBNetal96a}
H.B. Nielsen, H.H. Rugh and S.E. Rugh, ``Chaos and Scaling in
Classical Non-Abelian Gauge Fields'', Preprint LA-UR-96-1577,
chao-dyn/9605013 (May 1996).


\bibitem{Muller96a}
B. M\"{u}ller,  ``Study of Chaos and Scaling in Classical
$SU(2)$ Gauge Theory'', Preprint DUKE-TH-96-118,
chao-dyn/9607001 (July 1996).

\bibitem{HBNetal96b}
H.B. Nielsen, H.H. Rugh and S.E. Rugh, (in preparation).





\bibitem{hhrugh}
H. H. Rugh, ``Uniform Bounds on Lyapunov Exponents
in Lattice Gauge Theories'', (in preparation).
 

\bibitem{RughYMlic94}
S.E. Rugh, {\em Aspects of Chaos in the Fundamental Interactions.
Part I. Non-Abelian Gauge Fields}. (Part of)
Licentiate Thesis, The
Niels Bohr Institute. September 1994. 
Available upon request.
To appear (in a second revised edition) on the
e-print server.
 

\bibitem{Axenidesetal95}
M. Axenides, A. Johansen, H.B. Nielsen and O. T\"{o}rnkvist,
Nucl.Phys. {\bf B 474}, 3 (1996).

\bibitem{HotHadronicMatter95}
Cf. (e.g.) Sec. ``Thermalization - Entropy'', pp. 165 - 250 in
J. Letessier et al. (eds.), {\em Hot Hadronic Matter -
Theory and Experiment}. NATO ASI Series B: Physics Vol.
{\bf 346}. Plenum Press. New York. 1995.

\bibitem{FPU}
E. Fermi, J. Pasta and S. Ulam. Studies of nonlinear problems.
Report LA-1940, Los Alamos Scientific Laboratory, 1955.


\bibitem{MatinyanSavvidy}
S.G. Matinyan, G.K. Savvidy and N.G. Ter-Arutyuyan-Savvidy,
{\em Sov.Phys. JETP} {\bf 53}, 421 (1981);
G.K. Savvidy, {\em Nucl.Phys.} {\bf B 246}, 302 (1984).
S.G. Matinyan, {\em Sov.J.Part.Nucl.} {\bf 16},
226 (1985).


\bibitem{Chirikov}
B.V. Chirikov and D.L. Shepely\-anskii,
{\em JETP Lett.} {\bf 34}, 163 (1981);
{\em Sov. J. Nucl. Phys.} {\bf 36}, 908 (1982); 
E.S. Nikolaevskii and L.N. Shchur,
{\em Sov. Phys.JETP} {\bf 58}, 1 (1983).

\bibitem{Wellner}
M. Wellner, {\em Phys. Rev. Lett.} {\bf 68}, 1811 (1992);
{\em Phys. Rev.} {\bf E 50}, 780 (1994).


\bibitem{WilsonKogutSusskind} K. Wilson, Phys.Rev. D XX (1974).
J. Kogut and L. Susskind, {\em Phys. Rev.} {\bf D 11},
395 (1975).
 
\bibitem{Muller}
B. M\"uller and A. Trayanov, {\em Phys. Rev. Lett.} {\bf 68}, 3387 (1992).

 
\bibitem{Gong}
C. Gong, {\em Phys. Lett.} {\bf  B 298}, 257 (1993);
{\em Phys. Rev.} {\bf D 49}, 2642 (1994).
 
\bibitem{Biro}
T.S. Bir\'{o}, C. Gong, B. M\"{u}ller and A. Trayanov,
{\em Int. J. Mod. Phys.} {\bf C 5}, 113 (1994).



\bibitem{analogy}
See e.g. Chapt. 10 in R.P. Feynman and A.R. Hibbs,
{\em Quantum Mechanics and Path Integrals}. McGraw-Hill, Inc.
(1965); Chapt. 17-18 in H.J. Rothe, {\em Lattice Gauge
Theories}. World Scientific (1992).
 
\bibitem{Ambjorn} J.Ambj\o rn, T.Askgaard, H.Porter and
M.E.Shaposhnikov, {\em Nucl. Phys.} {\bf B353}, 346 (1991);
J. Ambj\o rn and A. Krasnitz, {\em Phys. Lett.} {\bf B 362},
97 (1995).

\bibitem{Kanaya}
K. Kanaya, ``Finite Temperature QCD on the Lattice'',
{\em Nucl.Phys.} {\bf B} (Proc.Suppl.) {\bf 47}, 144 (1996). 


\bibitem{Bodekeretal}
D. B\"{o}deker, L. McLerran and A. Smilga,
{\em Phys. Rev.} {\bf D 52}, 4675 (1995).
 
 
 
\bibitem{TBohr}
T. Bohr, ``Chaos and Turbulence'', in {\em Applications of
Statistical Mechanics and Field Theory to Condensed Matter},
edited by A.R. Bishop et al. Plenum Press. New York (1990);
W. van de Water and T. Bohr, {\em Chaos} {\bf 3}, 747 (1993).
 
 

\bibitem{Biroetal95} 
T.S. Bir\'{o}, C. Gong and B. M\"{u}ller,
Phys.Rev. {\bf D 52}, 1260 (1995).

\bibitem{Biroetal96}
T.S. Bir\'{o} and M.H. Thoma,
Phys.Rev. {\bf D 54}, 3465 (1996).

\bibitem{Heinzetal96}
U. Heinz, C.R. Hu, S. Leupold, S.G. Matinyan and
B. M\"{u}ller, ``Thermalization and Lyapunov exponents in the
Yang-Mills-Higgs Theory'', Preprint DUKE-TH-96-129,
hep-th/9608181 (August 1996).



 




\end{thebibliography}
\end{document}